**Presented at the first National Scientometrics Conference, 21-22 May, 2014, Isfahan University, Iran.**

# Analyzing readerships of International Iranian publications in Mendeley: an altmetrics study

By Zohreh Zahedi [1&2]

PhD candidate, Centre for science & Technology Studies (CWTS), Leiden University, The Netherlands[1]

Faculty member, department of & information Sciecnes, Persian Gulf University (PGU), Bushehr, Iran[2]

## Abstract

In this study, the presence and distribution of both Mendeley readerships (usage) and Web of Science (WoS) citations for the publications published in the 43 Iranian international journals indexed in JCR have been investigated. The aim was to determine the impact, visibility and use of the publications published by the Iranian international journals in Mendeley compared to their citation impact; furthermore, to explore if there is any relation between these two impact indicators (Mendeley readership counts and WoS citation counts) for these publications (i.e. the extent to which Mendeley readership counts correlate with citation indicators). The DoIs of the 1,884 publications[1] used to extract the readerships data from Mendeley REST API in February 2014 and citations data until end of 2013 calculated using CWTS in-house WoS database. SPSS (version 21) used to analyze the relationship between the readerships and citations for those publications. The Mendeley usage distribution both at the publication level (across publications years, fields and document types) and at the user level (across users' disciplines, academic status and countries) have been investigated. These information will help to understand the visibility and usage vs citation pattern and impact of Iranian scientific outputs. The findings indicate that 52% of those publications are saved in Mendeley; also, these publications on average have more readership per paper (RPP) (2.63) than citation per paper (CPP) (.49); also, the publications with at least one Mendeley readers exhibits more CPP (.49) compared to those that are not saved in Mendeley (i.e. with zero readership) (.39). This may indicate the benefit of saving document in Mendeley for these publications.

[1]. Out of 16,478 publications, 1,884 of them have DOIs out of which 1,389 were saved in Mendeley .

In terms of relations with citation, a weak spearman correlation (r = 0.179) has been found between Mendeley readership counts and citation indicators. This may suggest that users use these scientific outputs for reasons other than using and citing them in their scholarly practices (such as learning, teaching, self-awareness, etc.) which worth to explore in further research.

These findings also indicate that at the publication level, articles and reviews; publications from year 2013 and Medical and Engineering are among the most common document types, publication year and disciplines saved in Mendeley. Regarding the user level, Iranian international publications are mostly used by users from countries such as US, UK and Germany; PhDs as an academic status and disciplines such as Engineering, biological and medical sciences.

Mendeley readership data offers useful statistics and metadata about the impact and usage pattern of scientific outputs by different user types that are not available by citation databases; this information helps to formulate other types of impact than only scientific one and could be used as an informative and complimentary tool beside citation databases in interpreting impact of scholarly outputs. still many research needed to be done to dig more into these new metrics; their meaning and validity and reliability before its application in research evaluation.

# بررسی میزان استفاده از انتشارات انگلیسی زبان منتشر شده در مجلات بین المللی ایرانی در مندلی

زهره زاهدی

z.zahedi.2@cwts.leidenuniv.nl

دانشجوی دکتری، مرکز مطالعات علوم و فناوری، دانشگاه لایدن، هلند[2]
و
عضو هیئت علمی گروه علم اطلاعات و دانش شناسی، دانشگاه خلیج فارس بوشهر، ایران[3]

## چکیده

در این مطالعه، بصورت توصیفی، میزان حضور و توزیع میزان استفاده و میزان استناد مقالات منتشر شده در 43 مجله ایرانی نمایه شده در پایگاه گزارشهای استنادی مورد بررسی قرار گرفته است. توزیع میزان استفاده مقالات مذکور در مندلی در سطح نوع انتشار، سال انتشار و حوزه های موضوعی مقالات و همچنین در سطح کشور، رشته موضوعی و حرفه کاربران مندلی مطالعه شده است؛ همچنین رابطه (همبستگی) بین میزان استفاده و میزان استناد در سطح مقالات این مجلات مورد تجزیه و تحلیل قرار گرفته است. هدف اصلی بررسی میزان حضور و استفاده از مقالات بین المللی ایران در مندلی است تا بدینوسیله میزان روءیت پذیری آنها مشخص شود. یافته ها نشان داد که از نظر میزان، حدود نیمی از انتشارات جامعه آماری مورد بررسی در مندلی ذخیره شده اند و از نظر تاثیر استنادی، انتشاراتی که در مندلی ذخیره شده اند در مقایسه با آنهایی که مورد استفاده قرار نگرفته اند،رتبه استنادی بالاتری دارند؛ از لحاظ رابطه با استناد، رابطه همبستگی مثبت ولی ضعیف بین استناد و ذخیره مقالات در مندلی در بین انتشارات مورد بررسی وجود دارد.

---

2. Centre for Science & technology Studies (CWTS), Leiden University, The Netherlands
3. Knowledge & Information Science (KIS) Department, Persian Gulf University (PGU), Bushehr, Iran.

## مقدمه و بیان مسئله:

ایده آلتمتریکس[4]، اولین بار در سال 2010 توسط Priem و همکاران، بعنوان روشی نوین، غیرسنتی یا تکمیل کننده روشهای سنتی ارزیابی پژوهش در سنجش میزان تاثیر آثار علمی در محیط وب اجتماعی[5] مطرح شد (Priem, et. al, 2010). آلتمتریکس[6] یا دگرسنجه، ذکر آثار علمی در رسانه های وب اجتماعی نظیر فیس بوک[7]، توئیتر[8]، ویکیپدیا[9]، وبلاگ ها[10]، ابزارهای مدیریت استناد[11] نظیر مندلی[12]، رسانه های خبری[13] و غیره را در بر می گیرد. در واقع، محدودیتها و کاستیهای (Moed 2005; Bornmann & Daniel 2008) روشهای سنتی ارزیابی آثار پژوهشی (نظیر استناد و همتراز خوانی[14])، همواره محرک پژوهشگران در معرفی شاخصهایی (نظیر شاخص اچ[15]، وب سنجی و تحلیل پیوند[16] در محیط وب، میزان بارگیری[17] و فایلهای گزارش وب[18]) بوده است که نقش تکمیل کننده شاخصهای سنتی را ایفا نمایند (Shuai, Pepe & Bollen 2012; Thelwall 2008; Hirsch, 2005).

از زمان ظهور این ایده، گستره وسیعی از نرم افزارهای آلتمتریکس ایجاد شده اند که هر یک به تناسب دامنه و ویژگیهای خود قادرند تاثیرآنی[19] آثارعلمی را (در مقایسه با تاثیر استنادی که نیاز به زمان زیادی جهت محاسبه دارد) در رسانه های اجتماعی نمایش دهند؛ همین طور، آلتمتریکس بعنوان یک سنجه در سطح مقاله[20]، تاثیر یک مقاله را نه تنها در محدوده مقالات منتشر شده در مجلات و کنفرانسهای علمی بلکه در گستره وسیعی از منابع صرفنظر از فرمت انتشار نشان می دهد. بعنوان مثال، نشان دهنده تعداد دفعاتی است که یک اثر علمی در ویکیپدیا، وبلاگها، رسانه های خبری و سایر رسانه های وب 2.0 مطرح شده اند. البته، در هنگام تفسیر داده های آلتمتریکس بایستی محدودیتهایی کنونی آن نظیر نبود تعریفی مشخص و جامع، نبود شاخصی نرمال؛ در معرض دستکاری قرار گرفتن و همچنین

---

4 . Altmetrics or Article Level Metrics
5. Social Web
6. Altmetrics (Alternative metrics)
7. Facebook
8 . Twitter
9 . Wikipedia
10 . Weblogs
11 . Reference Management tools
12 . Mendeley
13 . News Media
14 . Peer Review
15 . H-index
16 . Webometrics and Link Analysis
17 . Downloads
18 . Web Log files
19. Real time impact
20. Article level Metrics (ALM)

نبود داده های استاندارد، مورد ملاحظه قرارداد( Wouters & Costas, 2012)؛همگی این موارد متاثر از جدید و نو مایه بودن این حوزه پژوهشی می باشد.

در مقابل، تحلیل استنادی و هم تراز خوانی همواره بعنوان رایج ترین و عمده ترین ابزارهای ارزشیابی کیفیت پژوهش، قدمتی دیرینه دارند و تا کنون بصورت گسترده در ارزشیابی پژوهش مورد استفاده قرار گرفته اند اما هریک به نوبه خود از نقاط قوت و ضعف خاص خود بی بهره نبوده اند (Moed, 2005) چرا که این دو رویکرد قادرند تنها قسمتی از تاثیر علمی آثار را نشان دهند ( Martin & Irvine, 1983)؛ بعبارت دیگر، یک سنجه به تنهایی بازتابنده تاثیر کامل آثار علمی نخواهد بود(Bollen et al., 2009). با توجه به این موارد، استفاده از روشهای تکمیل کننده (نظیر آلتمتریکس) در ارزشیابی پژوهش در کنار روشهای پیشین با اهمیت محسوب می گردد.

از میان ابزارهای آلتمتریکس، مندلی[21] یک نرم افزار مدیریت استناد رایگان است که از قابلیتهای زیادی جهت مدیریت، ذخیره، استناد واشتراک آثار پژوهشی برخوردار می باشد. علاوه بر این، مندلی پایگاه اطلاعاتی عظیمی از آثار علمی بشمار می رود بطوریکه حاوی بیشتر از 420 میلیون سند در رشته های مختلف علمی است که توسط بیش از 2.4 میلیون کاربر در آن ذخیره شده اند (Gunn, 2013).

ذخیره آثار علمی در مندلی تحت عنوان خواندن (Read) آن اثر معرفی شده است به این معنی که بصورت پیش فرض، افزودن هر اثر به کتابخانه شخصی ممکن است به مفهوم خواندن آن اثر توسط کاربر (در همان لحظه یا در آینده) و سپس استناد به آن در آثار علمی خود،اشاره داشته باشد. بدین ترتیب میزان کل استفاده (ذخیره) هر اثر توسط کاربران مختلف در مندلی با عنوان Readerships معرفی می گردد. یکی از خصیصه های قابل توجه مندلی، ارائه آماری از کل تعداد کاربرانی است که آثار علمی را در کتابخانه شخصی خود ذخیره کرده اند. علاوه بر این، آماری از توزیع کاربران بر حسب حرفه، کشور و رشته در سطح هر اثر علمی ذخیره شده در مندلی نیز وجود دارد؛ این اطلاعات تنها به 3% از کابران محدود می شود. به همین دلیل مندلی منبعی غنی از آلتمتریکس محسوب می شود چرا که علاوه بر ارائه میزان کل استفاده (ذخیره) از آثار علمی، اطلاعات فردی کاربران را نظیر توزیع استفاده در سطح حرفه، توزیع جغرافیایی و رشته در سطح هر اثر علمی نیز ارائه می نماید.

تحقیقات پیشین نشان داده اند که مندلی یکی از مهمترین ابزارهای آلتمتریکس (Li & Thelwall & Giustini 2012) و منبع مهمی حاوی آمار استفاده از آثار علمی حوزه موضوعی چند رشته ای (Zahedi, Costas & Wouters, 2014) بشمار می رود. عمده تحقیقات در این زمینه حول محور بررسی رابطه بین میزان استنادهای دریافتی در پایگا ه های استنادی در مقابل

21 . www.mendeley.com

میزان استفاده از مقالات در مندلی در حوزه های مختلف موضوعی انجام شده و درجات مختلفی از همبستگی (عمدتا همبستگی در حد کم و میانه) بین این دو مورد گزارش شده است (Henning, 2010; Priem et al., 2012; Li et al., 2012; Zahedi, Costas & Wouters, 2013; Schlögl et al., 2013; Thelwall et al., 2013; Haustein et al., 2013; Mohammadi & Thelwall, 2013; Costas, Zahedi & Wouters, 2014). سایر تحقیقات، امکان پذیری استفاده از آلتمتریکس بعنوان ابزاری جهت پیش بینی میزان استناد در آینده را مورد مطالعه قرار داده اند(Eysenbach, 2011; Waltman and Costas, 2013). نتیجه یک تحقیق در این زمینه نشان داده که بیشتر آثاری که در وبلاگها ذکر شده اند میزان استناد بیشتری نسبت به آثاری که ذکر نشده اند دریافت کرده اند (Shema et al., 2013). سایر مطالعات رابطه میان حرفه و کشور کاربران در مندلی را با میزان استنادهای دریافتی مجموعه ای از مقالات مورد مطالعه قرار داده اند( Zahedi, Costas & Wouters, 2013; Thelwall & Maflahi, in press).

مرور پیشینه ها نشان می دهد که با توجه به نو بودن مبحث آتمتریکس، مطالعات در این زمینه از سال 2010 میلادی به بعد انجام شده است. اکثر مطالعات با هدف بررسی رابطه بین استناد و آلتمتریکس پرداخته اند و نشان داده اند که عمدتاً بین این دو همبستگی کمی وجود دارد و نتیجه گرفته اند که ذکر آثار علمی در محیط وب اجتماعی و استناد به آثار علمی به نوعی بهم مرتبط هستند اما به دو فعالیت مختلف اشاره دارند که نیازمند پژوهش بیشتری است. همین طور، مندلی حاوی پوشش قابل توجهی از مقالات حوزه های مختلف موضوعی است و ذخیره مقالات در آن نشان دهنده نوعی از تاثیر این مقالات بر کاربران مختلف می باشد. همچنین، مرور پیشینه ها نشان داد که تا کنون مطالعه ای در زمینه بررسی حضور مجلات بین المللی ایران در مندلی انجام نشده است.

با توجه به اهمیت موضوع، در تحقیق حاضر میزان استفاده از مقالات انگلیسی زبان منتشر شده در مجلات بین المللی ایران در مندلی مورد بررسی قرار گرفته است. هدف این مطالعه، بررسی میزان حضور و ذخیره این مقالات در مندلی و نیز مقایسه بین میزان استفاده از آنها در آن و میزان استنادات دریافتی توسط این مقالات در پایگاه استنادی آی اس آی می باشد. بدین منظور، در این مقاله، توزیع میزان استفاده از مقالات در مندلی در سطح سال های انتشار، موضوعات و نوع انتشار مقالات و همچنین در سطح رشته، کشور و حرفه کاربران ارائه شده؛همچنین رابطه (میزان همبستگی)بین میزان ذخیره در مندلی و میزان استناد دریافتی در سطح مقالات مجلات مورد بررسی تجزیه و تحلیل قرار گرفته است.

در این راستا در این مقاله به سوالات پژوهشی زیر پاسخ داده شده است:

**1.** انتشارات مورد بررسی از چه میزان استناد و میزان استفاده در پایگاه استنادی آی اس آی و نرم افزار مدیریت استناد مندلی برخوردار هستند؟

**1.1** توزیع استناد و میزان استفاده از این انتشارات در پایگاه های فوق در سطح مقالات بر حسب موضوع، نوع انتشار و سال انتشار به چه صورت می باشد؟

**1.2** توزیع استناد و میزان استفاده از این انتشارات در پایگاه های فوق در سطح مقالات بر حسب حرفه، کشور و رشته کاربران مندلی به چه صورت می باشد؟

**2.** آیا رابطه همبستگی بین میزان استناد به این انتشارات در پایگاه استنادی آی اس آی و میزان استفاده از مقالات مورد مطالعه در نرم افزار مدیریت استناد مندلی وجود دارد؟

**روش‌شناسی:**

این مطالعه بصورت توصیفی و مکاشفه ای انجام شده است. جامعه آماری پژوهش حاضر مقالات منتشر شده در مجلات بین المللی ایرانی[22] (انگلیسی زبان) نمایه شده در پایگاه گزارشهای استنادی مجلات[23] است که از شناساگر دیجیتالی اشیاء[24] برخوردارهستند[25] سپس بر پایه شناساگر دیجیتالی اشیاء مقالات فوق، در بهمن ماه 1392 رابط برنامه نویسی نرم افزار[26] مندلی جهت گردآوری و استخراج دادهای آلتمتریکس استفاده شد؛ شاخص های کتابسنجی مقالات (حاوی تعداد کل استناد دریافتی تا سال 2013) با استفاده از نسخه سازمانی نمایه استنادی علوم در مرکز مطالعات علوم و فناوری دانشگاه لایدن[27] گردآوری و به فایل فوق افزوده شد تا بدین وسیله امکان مقایسه میزان استفاده از مقالات در مندلی و میزان استنادات دریافتی توسط مقالات مذکور فراهم شود. در نهایت از نرم افزار SPSS نسخه 21 جهت تجزیه و تحلیل داده ها استفاده شد تا بدین وسیله امکان مقایسه میزان استفاده و میزان استنادات دریافتی مقالات مذکور فراهم شود.

---

22. به پیوست نگاه کنید.
23 Journal Citation Reports (JCR)
24 Digital Object Identifier (DOI)
25 وجود شناساگر دیجیتالی اشیاء جهت استخراج داده های میزان استفاده از مقالات در مندلی حائز اهمیت می باشد
26 Application Programming Interface (API)
27 Centre for Science & technology Studies (CWTS), Leiden University

## یافته ها:

در این قسمت نتایج حاصل از تجزیه و تحلیل داده های فوق ارائه خواهد شد. در ابتدا، توزیع میزان استفاده مقالات مذکور در مندلی در سطح سال های انتشار، نوع انتشار، موضوعات مقالات و در سطح رشته، کشور و تخصص کاربران مقالات ارائه می شود، سپس میزان همبستگی بین میزان استفاده و استناد در سطح مقالات این مجلات مورد تجزیه و تحلیل قرار خواهند گرفت.

*توزیع کلی میزان استفاده و استناد در سطح مقالات*

جستجو در پایگاه گزارشهای استنادی مجلات نشان داد که در 43 مجله ایرانی نمایه شده در این پایگاه، 16487 مقاله منتشر شده است که از میان آنها، تنها 1884 از مقالات از شناساگر دیجیتالی اشیاء برخوردار هستند در صورتیکه 14603 از مقالات فاقد آن می باشند. از میان 1884 مقاله مورد بررسی، 1389 مقاله پس از جستجو در مندلی بازیابی شد و در تحلیل نهایی مورد استفاده قرار گرفت؛ از میان آنها 994(52.7%) مقاله حداقل یک بار در مندلی ذخیره شده اند در حالیکه 394 (20.9%) مقاله در مندلی اصلاً مورد استفاده قرار نگرفته اند. بررسی مقالاتی که حداقل یکبار در مندلی ذخیره شده اند نشان داد که این مقالات در بین سالهای 2011 تا 2103 منتشر شده اند؛ و بطور کلی 2624 بار در مندلی ذخیره شده اند و 489 استناد در پایگاه استنادی آی اس آی دریافت کرده اند. در مقایسه با میزان استناد، بطور میانگین، این مقالات از میزان استفاده بیشتری در مندلی (1.89) نسبت به میزان استناد(0.45)در پایگاه استنادی آی اس آی برخوردار هستند (جدول 1).

جدول 1. توزیع کلی میزان استفاده و استناد بر حسب کل انتشارات مورد بررسی

| Pubs | No | TCS | CPP | Total readerships | RPP |
|---|---|---|---|---|---|
| Without altm | 394 | 136 | .34 | 0 | |
| With alt | 994 | 489 | .49 | 2624 | 2.63 |
| Total Pubs | 1388 | 625 | .45 | | 1.89 |

### توزیع میزان استفاده و استناد بر حسب نوع انتشار

90% از کل انتشارات مورد بررسی از نظر نوع انتشار، مقاله مستند؛ بررسی توزیع میزان استفاده و میزان استناد دریافتی برحسب اشکال مختلف انتشار در انتشاراتی که حداقل یکبار در مندلی ذخیره شده نشان داد که 92.2% از کل مقالات مروری و 71.8% از کل مقالات جامعه آماری در مندلی ذخیره شده اند. در مقایسه با سایر انواع انتشارات، از نظر تعداد مقالات مروری بیشترین میزان ذخیره در مندلی دارند؛ از نظر میزان تاثیر، بطور میانگین، مقالات مروری از بیشترین میزان ذخیره در مندلی (2.57) برخوردار می باشند(جدول 2).

جدول 2. توزیع میزان استفاده و استناد بر حسب نوع انتشار

| Doc Type | Total Pubs | % | Total Citations | Pubs With alt | % | Total Citations | CPP | Total readerships | RPP |
|---|---|---|---|---|---|---|---|---|---|
| Articles | 1253 | 90.2 | 535 | 900 | 71.8 | 406 | .45 | 2316 | 2.57 |
| Non-citables | 43 | 3.1 | 0 | 26 | 60.5 | 0 | 0 | 67 | 2.57 |
| Letters | 37 | 2.7 | 1 | 16 | 43.2 | 1 | .06 | 21 | 1.31 |
| Reviews | 56 | 4.0 | 89 | 52 | 92.9 | 82 | 1.57 | 220 | 4.23 |
| Total | **1389** | 100 | 625 | 994 | 71.5% | 489 | .49 | 2624 | 2.63 |

### توزیع میزان استفاده و استناد بر حسب سال انتشار

باتوجه به جدول 3، علیرغم اینکه از نظر تعداد، انتشارات سال 2013 از بالاترین میزان پوشش در مندلی برخوردار هستند ( 84.5% از انتشارات منتشر شده در سال 2013 در جامعه مورد بررسی حداقل یکبار در مندلی ذخیره شده اند)، از نظر میزان استفاده، انتشارات سال 2012 بیشترین آمار ذخیره را در مندلی دارند ( این انتشارات 53.1% کل میزان استفاده را دریافت نموده اند).

جدول 3. توزیع میزان استفاده و استناد بر حسب سال انتشار

| Pub year | Total Pubs | % | Total Citations | Pubs With alt | % | Total Citations | CPP | Total readerships | RPP | % |
|---|---|---|---|---|---|---|---|---|---|---|
| **2011** | 216 | 15.6 | 186 | 168 | 77.8 | 172 | 1.02 | 689 | 4.1 | 26.3 |
| **2012** | 857 | 61.7 | 439 | 559 | 65.2 | 317 | .56 | 1394 | 2.49 | 53.1 |
| **2013** | 316 | 22.8 | 0 | 267 | 84.5 | 0 | 0 | 541 | 2.02 | 20.6 |
| **Total** | **1389** | 100 | **625** | **994** | | **489** | **.49** | **2624** | 2.63 | 100 |

*توزیع میزان استفاده و استناد بر حسب حوزه موضوعی انتشارات*

جدول 4، توزیع انتشارات کل جامعه آماری مورد بررسی را بر حسب موضوعات تحت پوشش آنها نشان می دهد. بطور کلی، پوشش موضوعی مقالات منتشر شده در مجلات مورد بررسی در 10 زمینه موضوعی در این جدول نشان داده شده است. انتشارات حوزه موضوعی مهندسی از بالاترین میزان پوشش در جامعه مورد بررسی برخوردار هستند (31.8% مقالات منتشر شده در مجلات مورد بررسی زمینه موضوعی مهندسی دارند). نکته قابل توجه این است که بطور متناسب، 100% مقالات حوزه پزشکی و 93.3% مقالات حوزه مهندسی محیط زیست حداقل یکبار در مندلی ذخیره شده اند؛ در حالیکه، مجلات حوزه شیمی از کمترین میزان ذخیره در مندلی برخوردار هستند. (بطور متناسب، 33.3% مقالات حوزه شیمی حداقل یکباردر مندلی ذخیره شده اند).

جدول 4. توزیع میزان استفاده و استناد بر حسب 10 حوزه موضوعی

| WOS Subject Categories | Total Pubs | % | Total CS | CPP | Pubs with alt | % | Total CS | CPP | Total RS | RPP |
|---|---|---|---|---|---|---|---|---|---|---|
| CHEMISTRY MULTIDISCIPLINARY | 180 | 13 | 46 | 0.3 | 60 | 33.3 | 21 | 0.4 | 102 | 1.7 |
| ENGINEERING ENVIRONMENTAL | 45 | 3.2 | 7 | 0.2 | 42 | 93.3 | 7 | 0.2 | 98 | 2.3 |
| ENGINEERING MULTIDISCIPLINARY | 442 | 31.8 | 269 | 0.6 | 346 | 78.3 | 237 | 0.7 | 1213 | 3.5 |
| ENVIRONMENTAL SCIENCES | 144 | 10.4 | 60 | 0.4 | 108 | 75.0 | 53 | 0.5 | 260 | 2.4 |
| GASTROENTEROLOGY & HEPATOLOGY | 157 | 11.3 | 111 | 0.7 | 119 | 75.8 | 81 | 0.7 | 291 | 2.4 |
| MEDICINE, GENERAL & INTERNAL | 29 | 2.1 | 1 | 0.0 | 29 | 100 | 1 | 0.0 | 71 | 2.4 |
| MICROBIOLOGY | 71 | 5.1 | 11 | 0.2 | 48 | 67.6 | 9 | 0.2 | 96 | 2.0 |
| PHARMACOLOGY & PHARMACY | 125 | 9 | 32 | 0.3 | 106 | 84.8 | 29 | 0.3 | 258 | 2.4 |
| POLYMER SCIENCE | 123 | 8.9 | 83 | 0.7 | 72 | 58.5 | 49 | 0.7 | 110 | 1.5 |
| RADIOLOGY, NUCLEAR MEDICINE & MEDICAL IMAGING | 73 | 5.3 | 5 | 0.1 | 64 | 87.7 | 2 | 0.0 | 125 | 2.0 |
| Total | **1389** | 100 | **625** | 0.4 | **994** | | **489** | 0.5 | **2624** | 2.6 |

*توزیع میزان استفاده از انتشارات مورد بررسی در مندلی بر حسب کشورهای مختلف کابران*

نمودار 1 توزیع میزان استفاده از انتشارات مورد بررسی را بر حسب کشورهای مختلف کاربران نمایش می دهد. کاربران از کشورهای مختلف (از 322 کشور مشخص و 2302 کشور نامشخص) 2624 بار انتشارات مورد نظر را در کتابخانه شخصی خود ذخیره کرده اند.در این نمودار، 10 کشور مشخص برتر از لحاظ میزان استفاده از انتشارات مورد بررسی در مندلی نشان داده شده است. کاربران ایالات متحده آمریکا، انگلستان و آلمان و پس از آن، ایران و هند کشورهای برتر از لحاظ میزان ذخیره و استفاده از این انتشارات می باشند. در واقع این مورد نشان دهنده رؤیت پذیری مقالات انگلیسی زبان منتشر شده در مجلات بین المللی ایران در بین کشورهای جهان است، هرچند که ممکن است این مقالات توسط کاربران ایرانی ساکن این کشورها مورد استفاده قرار گرفته باشند.

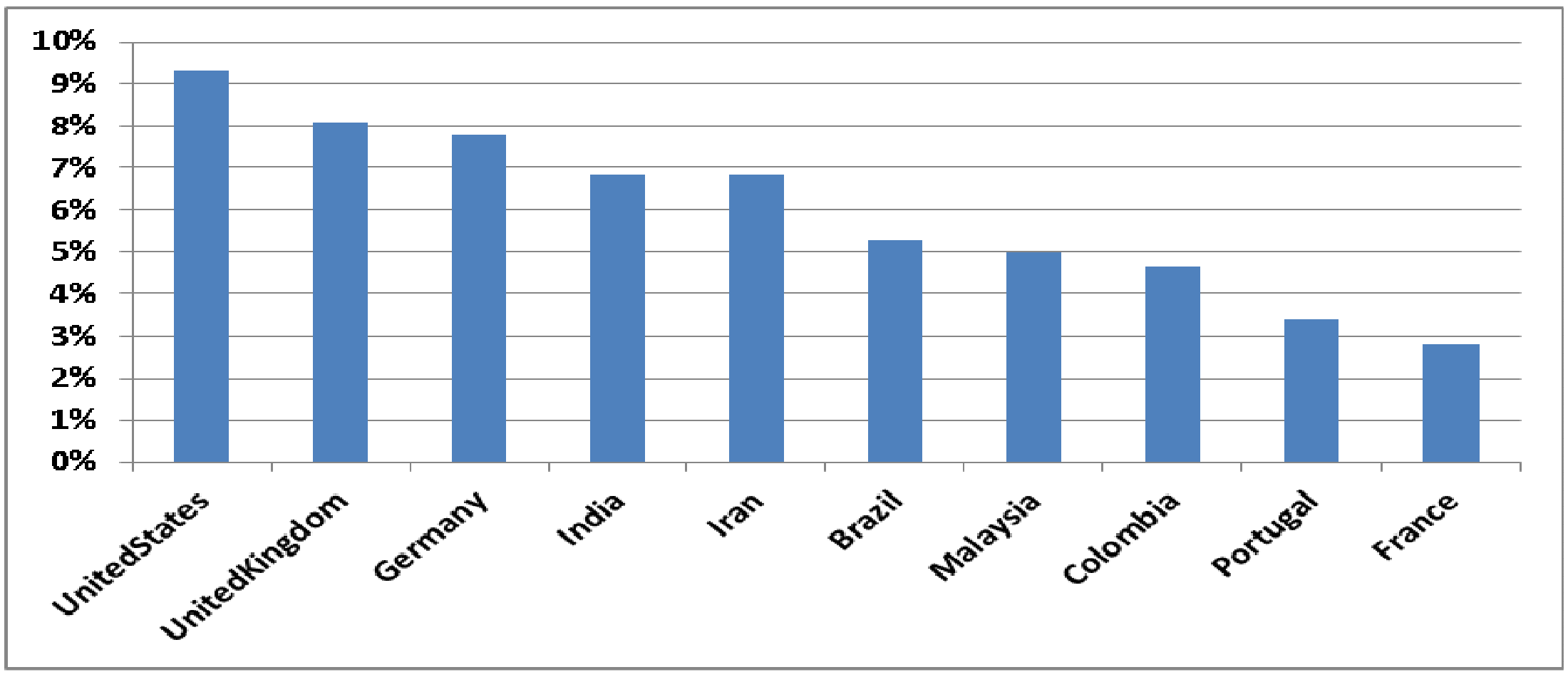

نمودار1. توزیع میزان استفاده از انتشارات مورد بررسی در مندلی بر حسب کشورهای مختلف کابران (10 کشور برتر)

*توزیع میزان استفاده از انتشارات مورد بررسی بر حسب حرفه کاربران در مندلی*

توزیع میزان استفاده از انتشارات مورد بررسی بر حسب حرفه کاربران مختلف در مندلی و بر حسب حوزه موضوعی انتشارات در نمودار 3 نمایش داده شده است. همانطور که مشخص است از میان انواع مختلف کاربران، تقریباً در همه حوزه های موضوعی، متداولترین کاربران در مندلی، دانشجویان و غیر متداولترین کاربران، کتابداران و مربیان هستند؛ تنها در حوزه موضوعی پزشکی، پزشکی هسته ای و تصویر برداری پزشکی متداولترین کاربران را پژوهشگران فوق دکتری تشکیل می دهد. این مورد می تواند گواه گرایش، تمایل و حتی آشنایی بیشتر دانشجویان

نسبت به سایر انواع کاربران به استفاده از ابزارهای وب 2.0 نظیر مندلی در فرایند انجام پژوهش باشد.

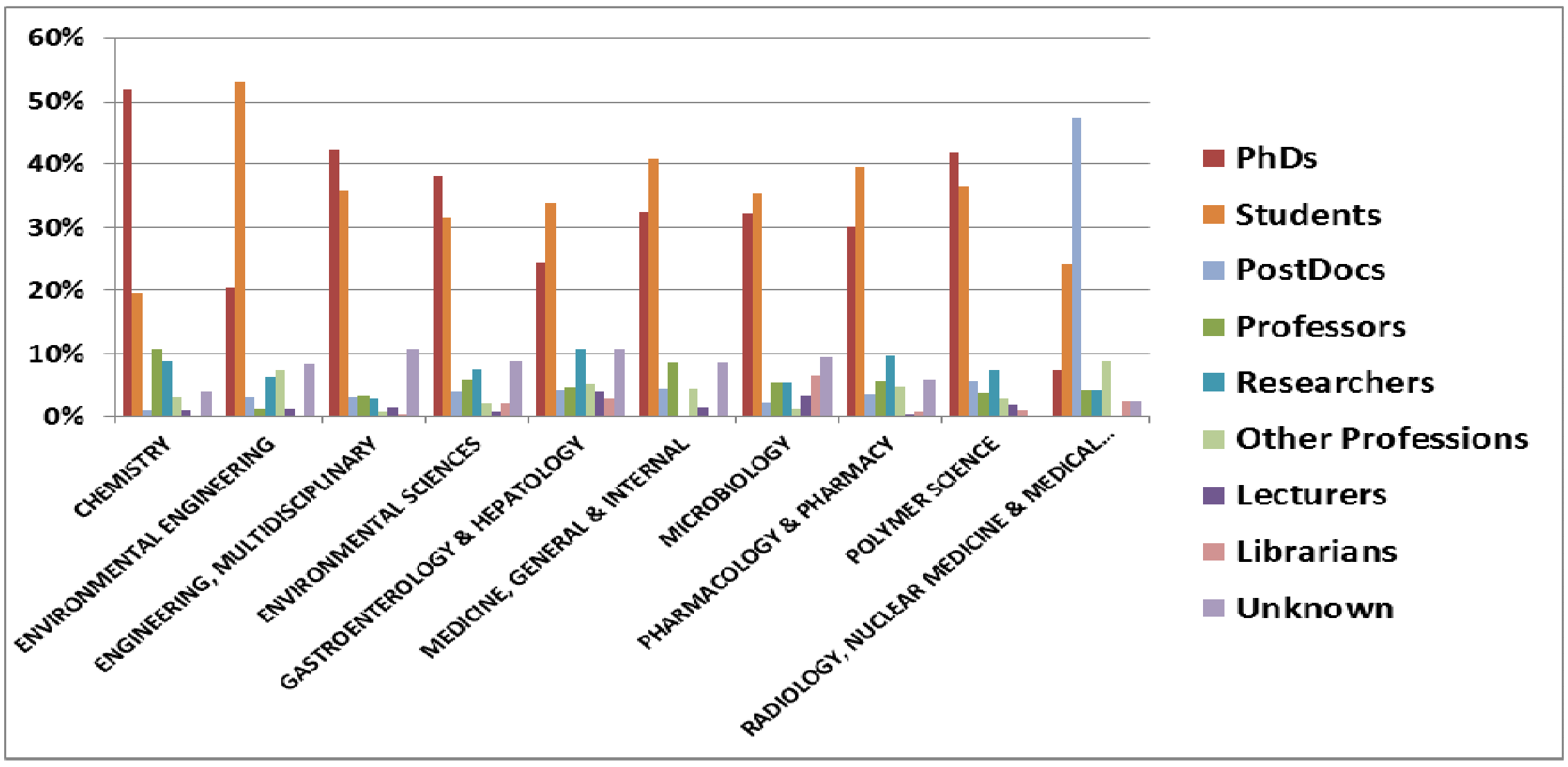

نمودار 2. توزیع میزان استفاده از انتشارات مورد بررسی بر حسب حرفه کاربران مختلف در مندلی

*توزیع میزان استفاده از انتشارات مورد بررسی بر حسب رشته تحصیلی- پژوهشی کاربران مندلی*

توزیع میزان استفاده از انتشارات مورد بررسی بر حسب رشته تحصیلی- پژوهشی کاربران مندلی در نمودار 3 نشان داده شده است. از کل 2624 بار میزان ذخیره انتشارات مورد بررسی در مندلی توسط کاربران مختلف، برای 2550 (%97.18) مورد، رشته کاربران مشخص و برای 74 (%2.82) مورد، رشته کاربران نامشخص بود. کاربران رشته های مهندسی، علوم زیستی، پزشکی و شیمی بالاترین میزان استفاده از مندلی را به خود اختصاص داده اند.

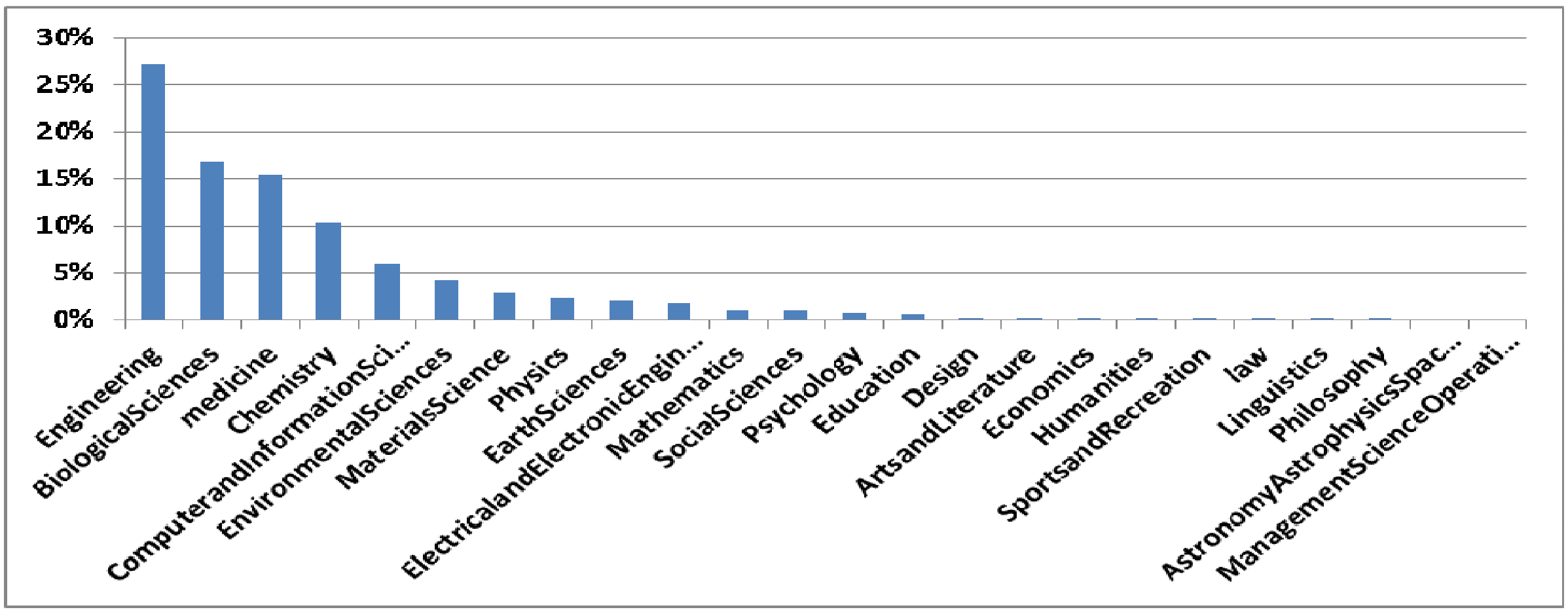

نمودار 3. توزیع میزان استفاده از انتشارات مورد بررسی بر حسب رشته تحصیلی- پژوهشی کاربران مندلی

*همبستگی بین میزان استناد و میزان استفاده انتشارات مورد بررسی در پایگاه استنادی آی اس آی و نرم افزار مدیریت استناد مندلی*

در این قسمت، رابطه بین میزان استناد و میزان استفاده از مقالات مورد بررسی مطالعه شده است؛ تا مشخص شود این دو مورد تا چه اندازه بهم مرتبط هستند. آزمون همبستگی اسپیرمن (با استفاده از روش بوت استرپینگ در بسته نرم افزاری علوم اجتماعی[28]) و با 95% فاصله اطمینان نشان داد که همبستگی مثبت و در حد ضعیف (r=.179) بین میزان استناد و میزان استفاده از انتشارات مورد بررسی در پایگاه استنادی آی اس آی و نرم افزار مدیریت استناد مندلی وجود دارد، بدین معنی که این دو متغیر بهم مرتبط اند اما با توجه به ضعیف بودن میزان همبستگی، به دو فعالیت متفاوت اشاره دارند(جدول 5).

جدول 5. رابطه همبستگی بین میزان استناد و میزان استفاده از انتشارات مورد بررسی

| Spearman's rho Correlation Coefficient | | Citations | readerships |
|---|---|---|---|
| Citations | | 1.000 | .179** |
| Readerships | | .179** | 1.000 |
| 95% Confidence Interval | Lower | 1.000 | .107 |
| | Upper | 1.000 | .253 |

**Correlation is significant at the 0.01 level (2-tailed)

همین طور رابطه همبستگی مثبت و ضعیفی بین مقالات استفاده شده توسط کاربران با حرفه های مختلف در مندلی با و میزان استناد به این مقالات مشاهده شد. از بین کاربران مختلف، مقلات ذخیره شده توسط دانشجویان، بالاترین میزان همبستگی را با میزان استناد دارد (r=.154)در مقایسه با سایر کاربران که همبستگی ضعیفی با استناد و با سایر کاربران دارند (جدول 6). همانطور که قبلاً هم ذکر شد در جامعه آماری مورد بررسی، بیشترین میزان کاربران مندلی را دانشجویان تشکیل می دهد.

---

28 . SPSS Bootstrapping method

جدول 6. رابطه همبستگی بین میزان استناد و میزان استفاده از انتشارات مورد بررسی بر حسب حرفه کاربران

| | Total Reader | PhD | Stud. | Res. | Post Docs | Prof. | Other Profess. | Librarian | Lecturer | unknown |
|---|---|---|---|---|---|---|---|---|---|---|
| **Citations** | .179** | .107** | .154** | -.018 | -.068 | .032 | -.029 | .036 | .092* | .185** |
| **lower** | .103 | .027 | .076 | -.085 | -.137 | -.045 | -.093 | -.048 | .015 | .105 |
| **upper** | .250 | .179 | .224 | .048 | .002 | .108 | .046 | .112 | .171 | .261 |
| **Total Readers** | | .582** | .570** | .135** | .028 | .074* | .066 | .085* | .121** | .592** |
| **lower** | | .527 | .510 | .064 | -.037 | .002 | -.005 | .005 | .054 | .546 |
| **upper** | | .635 | .624 | .202 | .101 | .142 | .133 | .160 | .190 | .638 |
| **PhDs** | | | .052 | -.093* | -.204** | -.070 | -.164** | -.064 | .030 | .329** |
| **lower** | | | -.027 | -.160 | -.267 | -.141 | -.222 | -.130 | -.043 | .254 |
| **upper** | | | .131 | -.012 | -.125 | .000 | -.103 | .004 | .102 | .401 |
| **Students** | | | | -.109** | -.146** | -.125** | -.086* | -.024 | -.050 | .441** |
| **lower** | | | | -.181 | -.213 | -.196 | -.151 | -.096 | -.125 | .372 |
| **upper** | | | | -.042 | -.077 | -.048 | -.017 | .056 | .030 | .503 |
| **Researchers** | | | | | -.038 | -.044 | .021 | -.028 | .010 | .054 |
| **lower** | | | | | -.100 | -.102 | -.060 | -.067 | -.058 | -.023 |
| **upper** | | | | | .025 | .023 | .107 | .049 | .098 | .143 |
| **PostDocs** | | | | | | -.083* | .068 | -.002 | -.030 | -.017 |
| **lower** | | | | | | -.132 | -.015 | -.062 | -.084 | -.080 |
| **upper** | | | | | | -.026 | .169 | .087 | .041 | .062 |
| **Professors** | | | | | | | -.089* | .006 | -.051 | -.030 |
| **lower** | | | | | | | -.106 | -.055 | -.083 | -.099 |
| **upper** | | | | | | | -.072 | .097 | .003 | .040 |
| **Other Professionals** | | | | | | | | .129** | -.016 | .082* |
| **lower** | | | | | | | | .002 | -.055 | -.017 |
| **upper** | | | | | | | | .281 | .060 | .174 |
| **Librarians** | | | | | | | | | -.030 | .135** |
| **lower** | | | | | | | | | -.039 | .015 |
| **upper** | | | | | | | | | -.021 | .241 |
| **Lecturers** | | | | | | | | | | .117** |
| **lower** | | | | | | | | | | .019 |
| **upper** | | | | | | | | | | .223 |

**Correlation is significant at the 0.01 and 0.05 levels (2-tailed)

## بحث و نتیجه‌گیری:

مندلی ابزار قدرتمندی از نظر برخورداری از داده های آلتمتریکس قلمداد می شود بطوریکه ذخیره مقالات در آن می تواند نشان دهنده نوعی از تاثیر این مقالات بر کاربران باشد. نتایج این پژوهش نشان داد که مجلات ایرانی بین المللی پوشش تقریبا خوبی در مندلی دارند بطوریکه حدود نیمی از انتشارات آنها در مندلی ذخیره شده اند که نشان دهنده مورد استفاده قرار گرفتن و رویت پذیری مقالات مورد بررسی توسط کاربران مختلف را در مندلی است و از این نظر قابل توجه می باشد. البته نباید محدودیت کنونی مندلی در عدم ارائه آمار کامل از کاربران را از نظر دور داشت.

یافته های این مطالعه نشان داد که انتشاراتی که در مندلی ذخیره شده اند در مقایسه با آنهایی که استفاده نشده اند،رتبه استنادی بالاتری دارند. از نظر نوع انتشار، سال انتشار و حوزه موضوعی، به ترتیب، مقالات،انتشارات سال 2012 و انتشارات حوزه پزشکی از بیشترین میزان ذخیره در مندلی برخوردار می باشند. همچنین، بر اساس کاربران، تقریباً در همه حوزه های موضوعی، متداولترین کاربران در مندلی را دانشجویان تشکیل می دهد، این مورد تایید کننده نتیجه تحقیق قبلی در زمینه کاربران مندلی می باشد (Zahedi, Costas & Wouters; 2013)؛ بر حسب رشته، کاربران رشته مهندسی بالاترین میزان استفاده از مندلی را به خود اختصاص داده اند و از نظر کشور بیشترین کاربران از ایالات متحده آمریکا هستند. مطالعه استنادی و وب سنجی مجلات ایرانی بین المللی مشخص کرده اند که این مجلات از رویت پذیری کمی در سطح بین المللی برخوردارند( Zahedi, 2008; Davarpana & Behrouzfar, 2009). از نظر رابطه با استناد، بطور کلی، همبستگی مثبت و ضعیف بین استناد و ذخیره مقالات در مندلی در بین انتشارات مورد بررسی وجود دارد. این نتیجه تایید کننده یافته های تحقیقات پیشین در این زمینه می باشد ( Zahedi, Costas & Wouters, 2013; Haustein et al., 2013; etc.

بطور کلی، داده های مندلی در مورد میزان استفاده از آثار علمی منبع جدیدی برای بررسی عمیق تر تاثیر آثار علمی در اختیار محققان

این حوزه قرار داده است که می تواند بعنوان شاخص جدیدی در سنجش اثر گذاری آثار علمی توسط گستره وسیعی از کاربران ارائه نماید، همین طور، در کنار سایر روشهای سنجش تاثیر آثار علمی بعنوان یک سنجه تکمیل کننده معرفی گردد. از طرف دیگر، با توجه به نومایه بودن حوزه پژوهشی آلتمتریکس، پژوهشهای این حوزه هنوز در ابتدای راه است به همین دلیل هنوز بصورت دقیق مفهوم آلتمتریکس و دلیل اصلی استفاده از ابزارهای وب 2.0 توسط کاربران مشخص نیست: سوالاتی نظیر اینکه "آیا کاربران در استفاده از این ابزارها از انگیزه مشابه یا متفاوتی نسبت به زمانی که به آثار پژوهشی استناد می کنند برخوردارند"، نیازمند پژوهش بیشتر می باشد. علاوه بر این، پژوهشگران معتقدند که علیرغم اینکه این ابزارها قادرند تاثیر آنی آثار علمی در رسانه های اجتماعی را نشان دهند، در حال حاضر، تنها بعنوان یک ابزار برای خود ارزیابی توسط پژوهشگران مفید هستند، و هنوز جهت ارزشیابی نظام مند پژوهش مناسب نیستند. استانداردسازی ابزار و داده ها و کیفیت داده ها از ضروریات است که قبل از بکارگیری در فرایند ارزیابی پژوهش باید مورد توجه قرار گیرد (Wouters & Costas, 2012).

**منابع:**

Bollen, J., Van de Sompel, H., Hagberg, A., & Chute, R. (2009). A principal component analysis of 39 scientific impact measures. PLoS ONE, 4(6), e6022. doi:10.1371/journal.pone.0006022.

Bornmann, L., & Daniel, H.-D. (2008). What do citation counts measure? A review of studies on citing behavior. Journal of Documentation, 64(1), 45–80.

Davarpana, M. & Behrouzfar, H.(2009). International visibility of Iranian ISI journals: A citation study. Aslib Proceedings, Vol. 61(4): 407 – 419.

Eysenbach, G. (2011). Can tweets predict citations? Metrics of social impact based on twitter and correlation with traditional metrics of scientific impact. Journal of Medical Internet Research, 13(4), e123.

Gunn, W. (2013). Social Signals Reflect Academic Impact: What it Means When a Scholar Adds a Paper to Mendeley. Information Standards Quarterly. 25(2): 33-39.

Haustein, S., Peters, I., Bar-Ilan, J., Priem, J., Shema, H., & Terliesner, J. (2013). Coverage and adoption of altmetrics sources in the bibliometric community. In 14th International Society of Scientometrics and Informetrics Conference (pp. 1–12). Digital Libraries. Retrieved from http://arxiv.org/abs/1304.7300.

Hirsch, J. E. (2005), An index to quantify an individual's scientific research output. Proceedings of the National Academy of Sciences of the United States of America, 102 : 16569–16572.

Li, X., Thelwall, M., & Giustini, D. (2012). Validating online reference managers for scholarly impact measurement. Scientometrics, 91(2), 461–471.

Martin, B. R., & Irvine, J. (1983). Assessing basic research: Some partial indicators of scientific progress in radio astronomy. Research Policy, 12, 61–90.

Moed, H.F. (2005). Citation analysis in research evaluation. Berlin/ Heidelberg/New York: Springer.

Mohammadi, E. & Thelwall, M. (2013). Assessing the Mendeley Readership of Social Sciences and Humanities Research. In Proceedings of the 14th International Society of Scientometrics and Informetrics Conference (Vol. 1, pp.200-214).

Priem, J., Taraborelli, D., Groth, P., and Neylon, C. (2010). Altmetrics: a manifesto. Retrieved December 10, 2012 from: http://altmetrics.org/manifesto/

Schlo¨gl, C., Gorraiz, J., Gumpenberger, C., Jack, K., & Kraker, P. (2013). Download vs. citation vs. readership data: the case of an information systems journals. In J. Gorraiz, E. Schiebel, C. Gumpenberger, M. Hörlesberger, & H. Moed (Eds.), Proceedings of the 14th International Society of Scientometrics and Informetrics Conference, Vienna, Austria (pp. 626-634). Wien: Facultas Verlags und Buchhandels AG.

Shema, H., Bar-Ilan, J., & Thelwall, M. (2013). Do blog citations correlate with a higher number of future citations ? Research blogs as a potential source for alternative metrics. In J. Gorraiz, E. Schiebel, C. Gumpenberger, M. Hörlesberger, & H. Moed (Eds.), Proceedings of the 14th International Society of Scientometrics and Informetrics Conference, Vienna, Austria (pp. 604–611). Wien: Facultas Verlags und Buchhandels AG.

Shuai, X, Pepe, A, Bollen, J. (2012). How the scientific community reacts to newly submitted preprints: article downloads Twitter mentions, and citations. Retrieved 25 December 2012 from: ArXiv: 1202.2461v1201

Thelwall, M. (2008). Bibliometrics to Webometrics, Journal of Information Science, 34(4), 605-621.

Thelwall, M., Haustein, S., Larivière, V., & Sugimoto, C. R. (2013). Do altmetrics work? Twitter and ten other social web services. PLoS ONE, 8(5), e64841.

Thelwall, M. & Maflahi, N. (in press). Are scholarly articles disproportionately read in their own country? An analysis of Mendeley readers. Journal of the American Society for Information Science and Technology.

Waltman, L., & Costas, R. (2013). F1000 Recommendations as a potential new data source for research evaluation: a comparison with citations. Journal of the Association for Information Science and Technology. doi: 10.1002/asi.23040.

Wouters, P., Costas, R. (2012). Users, narcissism and control: Tracking the impact of scholarly publications in the 21st century. Utrecht: SURF foundation. Retrieved September 20, 2012 from: http://www.surffoundation.nl/nl/publicaties/Documents/Users%20narcissism%20and%20control.pdf

Zahedi, Z. (2008). Visibility of Iranian journals websites: a webometrics study. In Kretschmer, H., Havemann, F. (Eds), Proceedings of WIS 2008, Berlin, Fourth International Conference on Webometrics, Infometrics and Scientometrics and Ninth COLLNET Meeting, Humboldt-Universitat zu Berlin, Institute for Library and Information Science (IBI), available at: http://creativecommons.org/liceness/by/2.0/ (accessed 15 October 2008).

Zahedi, Z., Costas, R. & Paul Wouters. (2013). How well developed are Altmetrics? Cross disciplinary analysis of the presence of ‘alternative metrics’ in scientific publications. In Proceedings of the 14th International Society of Scientometrics and Informetrics Conference (Vol. 1, pp. 876–884)

Zahedi, Z., Costas, R., & Wouters, P. (2013). What is the impact of the publications read by the different Mendeley users? Could they help to identify alternative types of impact? Presentation held at the PLoS ALM Workshop 2013 in San Francisco, http://lanyrd.com/2013/alm13/scrdpk/.

Zahedi, Z., Costas, R. & Paul Wouters. (2014). How well developed are Altmetrics? Cross disciplinary analysis of the presence of ‘alternative metrics’ in scientific publications. Scientometrics, DOI:10.1007/s11192-014-1264-0.

**پیوست:**

فهرست 43 مجله نمایه شده در پایگاه گزارشهای استنادی مجلات

| No. | Journal Title | Abbr. | ISSN |
|---|---|---|---|
| 1 | ARCHIVES OF IRANIAN MEDICINE | RCH IRAN MED | 1029-2977 |
| 2 | BANACH JOURNAL OF MATHEMATICAL ANALYSIS | BANACH J MATH ANAL | 1735-8787 |
| 3 | BULLETIN OF THE IRANIAN MATHEMATICAL SOCIETY | B IRAN MATH SOC | 1735-8515 |
| 4 | CELL JOURNAL | CELL J | 2228-5806 |
| 5 | DARU-JOURNAL OF FACULTY OF PHARMACY | DARU | 1560-8115 |
| 6 | HEPATITIS MONTHLY | HEPAT MON | 1735-143X |
| 7 | INTERNATIONAL JOURNAL OF CIVIL ENGINEERING | INT J CIV ENG | 1735-0522 |
| 8 | INTERNATIONAL JOURNAL OF ENVIRONMENTAL RESEARCH | INT J ENVIRON RES | 1735-6865 |
| 9 | INTERNATIONAL JOURNAL OF ENVIRONMENTAL SCIENCE AND TECHNOLOGY | INT J ENVIRON SCI TE | 1735-1472 |
| 10 | INTERNATIONAL JOURNAL OF FERTILITY & STERILITY | INT J FERTIL STERIL | 2008-076X |
| 11 | INTERNATIONAL JOURNAL OF PLANT PRODUCTION | INT J PLANT PROD | 1735-6814 |
| 12 | IRANIAN JOURNAL OF ALLERGY ASTHMA AND IMMUNOLOGY | IRAN J ALLERGY ASTHM | 1735-1502 |
| 13 | IRANIAN JOURNAL OF ARTHROPOD-BORNE DISEASES | IRAN J ARTHROPOD-BOR | 1735-7179 |
| 14 | IRANIAN JOURNAL OF BASIC MEDICAL SCIENCES | IRAN J BASIC MED SCI | 2008-3866 |
| 15 | IRANIAN JOURNAL OF CHEMISTRY & CHEMICAL ENGINEERING-INTERNATIONAL ENGLISH EDITION | IRAN J CHEM CHEM ENG | 1021-9986 |
| 16 | IRANIAN JOURNAL OF ENVIRONMENTAL HEALTH SCIENCE & ENGINEERING | IRAN J ENVIRON HEALT | 1735-1979 |
| 17 | IRANIAN JOURNAL OF FISHERIES SCIENCES | IRAN J FISH SCI | 1562-2916 |
| 18 | IRANIAN JOURNAL OF IMMUNOLOGY | IRAN J IMMUNOL | 1735-1383 |
| 19 | IRANIAN JOURNAL OF KIDNEY DISEASES | IRAN J KIDNEY DIS | 1735-8582 |
| 20 | IRANIAN JOURNAL OF OPHTHALMOLOGY | IRAN J OPHTHALMOL | 1735-4153 |
| 21 | IRANIAN JOURNAL OF PARASITOLOGY | IRAN J PARASITOL | 1735-7020 |
| 22 | IRANIAN JOURNAL OF PEDIATRICS | IRAN J PEDIATR | 2008-2142 |
| 23 | IRANIAN JOURNAL OF PHARMACEUTICAL RESEARCH | IRAN J PHARM RES | 1735-0328 |
| 24 | IRANIAN JOURNAL OF PUBLIC HEALTH | IRAN J PUBLIC HEALTH | 2251-6085 |
| 25 | IRANIAN JOURNAL OF RADIATION RESEARCH | IRAN J RADIAT RES | 1728-4554 |
| 26 | IRANIAN JOURNAL OF RADIOLOGY | IRAN J RADIOL | 1735-1065 |
| 27 | IRANIAN JOURNAL OF REPRODUCTIVE MEDICINE | IRAN J REPROD MED | 1680-6433 |
| 28 | IRANIAN JOURNAL OF SCIENCE AND TECHNOLOGY TRANSACTION A-SCIENCE | IRAN J SCI TECHNOL A | 1028-6276 |
| 29 | IRANIAN JOURNAL OF SCIENCE AND | IRAN J SCI TECHNOL B | 1028-6284 |
| | TECHNOLOGY TRANSACTION B-ENGINEERING | | |
| 30 | IRANIAN JOURNAL OF SCIENCE AND TECHNOLOGY-TRANSACTIONS OF ELECTRICAL ENGINEERING | IJST-T ELECTR ENG | 2228-6179 |
| 31 | IRANIAN JOURNAL OF SCIENCE AND TECHNOLOGY-TRANSACTIONS OF MECHANICAL ENGINEERING | IJST-T MECH ENG | 2228-6187 |
| 32 | IRANIAN JOURNAL OF VETERINARY RESEARCH | IRAN J VET RES | 1728-1997 |
| 33 | IRANIAN POLYMER JOURNAL | IRAN POLYM J | 1026-1265 |
| 34 | IRANIAN RED CRESCENT MEDICAL JOURNAL | IRAN RED CRESCENT ME | 1561-4395 |
| 35 | JOURNAL OF AGRICULTURAL SCIENCE AND TECHNOLOGY | J AGR SCI TECH-IRAN | 1680-7073 |
| 36 | JOURNAL OF APPLIED FLUID MECHANICS | J APPL FLUID MECH | 1735-3572 |
| 37 | JOURNAL OF ARTHROPOD-BORNE DISEASES | J ARTHROPOD-BORNE DI | 1735-7179 |
| 38 | JOURNAL OF RESEARCH IN MEDICAL SCIENCES | J RES MED SCI | 1735-1995 |
| 39 | JOURNAL OF THE IRANIAN CHEMICAL SOCIETY | J IRAN CHEM SOC | 1735-207X |
| 40 | JUNDISHAPUR JOURNAL OF MICROBIOLOGY | JUNDISHAPUR J MICROB | 2008-3645 |
| 41 | SCIENTIA IRANICA | SCI IRAN | 1026-3098 |
| 42 | UROLOGY JOURNAL | UROL J | 1735-1308 |
| 43 | YAKHTEH | YAKHTEH | 1561-4921 |